\documentclass[8.5pt,twoside,twocolumn]{article}
\usepackage[super,sort&compress,comma]{natbib} 
\usepackage{mhchem}
\usepackage{times,mathptm}
\usepackage{sectsty}
\usepackage{balance} 

\usepackage{graphicx} 
\usepackage{lastpage}
\usepackage[format=plain,justification=raggedright,singlelinecheck=false,font=small,labelfont=bf,labelsep=space]{caption} 
\usepackage{fancyhdr}

\begin{document}

\thispagestyle{plain}
\setcounter{secnumdepth}{5}

\makeatletter 
\def\subsubsection{\@startsection{subsubsection}{3}{10pt}{-1.25ex plus -1ex minus -.1ex}{0ex plus 0ex}{\normalsize\bf}} 
\def\paragraph{\@startsection{paragraph}{4}{10pt}{-1.25ex plus -1ex minus -.1ex}{0ex plus 0ex}{\normalsize\textit}} 
\renewcommand\@biblabel[1]{#1}            
\renewcommand\@makefntext[1]%
{\noindent\makebox[0pt][r]{\@thefnmark\,}#1}
\makeatother 
\renewcommand{\figurename}{\small{Fig.}~}
\sectionfont{\large}
\subsectionfont{\normalsize} 

\renewcommand{\headrulewidth}{1pt} 
\renewcommand{\footrulewidth}{1pt}
\setlength{\arrayrulewidth}{1pt}
\setlength{\columnsep}{6.5mm}
\setlength\bibsep{1pt}

\twocolumn[
  \begin{@twocolumnfalse}
\noindent\LARGE{\textbf{Scaling between Structural Relaxation and Particle Caging in a Model Colloidal Gel}}
\vspace{0.6cm}

\noindent\large{\textbf{C. De Michele,$^{\ast}$\textit{$^{a}$} E. Del Gado,
\textit{$^{b}$} and
D. Leporini\textit{$^{c}$}}}\vspace{0.5cm}


\vspace{0.6cm}

\noindent \normalsize{
In polymers melts and supercooled liquids, the glassy dynamics is characterized by the rattling of monomers or particles in the cage formed by their neighbors. Recently, 
a direct correlation in such systems, described by a universal scaling form, has been established  between the rattling amplitude and the structural relaxation time.  
In this paper we analyze the glassy dynamics emerging from the formation of a persistent network in a model colloidal gel at very low density.
The structural relaxation time of the gel network is compared with the mean squared
displacement at short times, corresponding to the localization length associated to
the presence of energetic bonds. Interestingly, we find that the same type of scaling as for
the dense glassy systems holds. Our findings well elucidate the strong coupling between 
the cooperative rearrangements of the gel network and the single particle localization 
in the structure. Our results further indicate that the scaling captures indeed fundamental 
physical elements of glassy dynamics.


 
}
\vspace{0.5cm}
 \end{@twocolumnfalse}
  ]

\section{Introduction}

Understanding the extraordinary slow-down that accompanies systems with structural arrest like glass-forming systems  \cite{Angell95,DebeStilli2001} and soft matter\cite{frenkel,pham,vlassopoulos, luca_jpcm05,SollichEtAl09,EZJPCM07,LikosPhysRep01} is a major scientific challenge. 
Crowding and caging effects play major roles in the glass transition (GT) of dense systems and lead to the strong localization of the particles
in the cage formed by their neighbors: 
this is typically apparent in the time dependence of the particle mean square displacement (MSD) as a plateau-like regime.

The value of the MSD plateau $\langle u^2\rangle$ yields the amplitude of the rattling motion 
inside the cage and hence the corresponding localization length.

Because of the extreme time-scale separation between the rattling
motion ($ \sim 10^{-12}\;\textrm{s}$) and the structural relaxation  ($
\tau_\alpha \sim 10^2\textrm{s}$ at GT), one would expect their complete
independence. 
\footnotetext{\textit{$^{a}$~Dipartimento di Fisica, ``Sapienza''
Universit\`a di Roma, P.le A. Moro 2, Roma, I-00185, Italy, Fax:+3906463158, Tel: +390649913524, E-mail: cristiano.demichele@roma1.infn.it}}
\footnotetext{\textit{$^{b}$~ETH Z\"urich, Department of Civil Engineering, Microstructure and Rheology, CH-8093, Z\"urich, Switzerland, E-mail: delgado@ethz.ch }}
\footnotetext{\textit{$^{c}$~IPCF-CNR and Dipartimento di Fisica ``E. Fermi'', Universit\`a di Pisa, Largo B.\@Pontecorvo 3, Pisa, I-56127, Italy, E-mail: leporini@df.unipi.it}}

Nonetheless, there are several hints of the presence of correlations and several authors have investigated them\cite{TobolskyEtAl43, Angell68,Nemilov68,HallWoly87,BuchZorn92, Ngai00,StarrEtAl02,NoviSoko03,ScopignoEtAl03,NoviSoko04,Ngai04,Dyre04,Dyre06,Harrowell_NP08, Johari06,OurNatPhys,OttochianEtAl09,OttochianLepoJNCS10,OttochianLepoPhilMag10,OttochianLepoPhilMag08}. 
In particular, it has been recently shown that the structural relaxation time $\tau_\alpha$ and the 
rattling amplitude $\langle u^2\rangle^{1/2}$ of several numerical models, 
including linear polymers, mixtures, prototypical glassformers like $SiO_2$ and o-terphenyl (OTP), 
and one icosahedral glassformer \cite{OurNatPhys,OttochianEtAl09,OttochianLepoJNCS10}, can be related in a unique scaling form.

Remarkably, the same resulting master curve well fits to the
experimental data from van der Waals and associating liquids, polymers, metallic glasses, ionic 
liquids and network glassformers over many decades in time
\cite{OurNatPhys,OttochianEtAl09,OttochianLepoJNCS10,OttochianLepoPhilMag10}.
These results well elucidate how, in an impressively large class of systems, the glassy structural arrest 
corresponds to the onset of a strong coupling between the overall relaxation, characterized by 
cooperative and heterogeneous processes, and the average localization at the level of a single particle, strongly suggesting that this is a universal, fundamental feature of this type of dynamics.  


Structural arrest and glassy dynamics can be observed also in very dilute particle suspensions when 
gelation occurs and displays significant hints of caging effects even at rather low volume fractions 
\cite{edg_pre04,puertas04,puertas05,DeMicheleSticky06,DeMicheleSilica06,JCPnmax06,edgkob_epl05,edgkob_prl07,edgkob_sm10,fofficriJCP05}. 
In these cases, differently from the dense glass-forming systems just mentioned, 
most of the times the caging has been associated to the particle bonding, rather than to the 
role of excluded volume interactions \cite{ema_epl03,puertas02,af_jstat08}. 
In these systems, particles get bonded into an interconnected network structure which is
responsible for the onset of cooperative, slow dynamics and eventually structural arrest,
i.e. {\it gelation}. It has been recently shown, in the numerical study of a model colloidal gel, 
that the gel network induces the same type of strong coupling in particle motion typically 
observed in dense glassy systems and that the glassy dynamics directly arises from the 
cooperative processes induced by the network \cite{edgkob_sm10}. A distinctive feature of this type of
systems is the presence of different localization processes, over different length scales, leading 
to a somewhat more complex scenario for structural arrest. 
Nevertheless, particle caging is also observed, albeit much weaker than in dense systems, and again over time scales which are well separated from the ones of 
structural relaxation.
Intrigued by the similarities and differences in the glassy dynamics of dense glasses and low 
volume fraction gels, we have investigated the presence and nature of correlations between 
structural relaxation and particle localization in a model colloidal gel.
In spite of the deep differences in the caging mechanism and onset of slow dynamics, we find that the same universal scaling, already found in dense glassy systems, between the structural 
relaxation time $\tau_\alpha$ and $\langle u^2\rangle$ holds
\cite{OurNatPhys,OttochianEtAl09,OttochianLepoJNCS10,OttochianLepoPhilMag10}.
In our view this finding points to the presence of a complex {\it feedback} between 
the overall slow structural relaxation and the single particle localization in dilute gels. 
At the same time, it further supports the idea that this scaling form captures 
a fundamental, universal feature of glassy structural arrest.

\begin{figure}[h]
\centering
  \includegraphics[width=\linewidth]{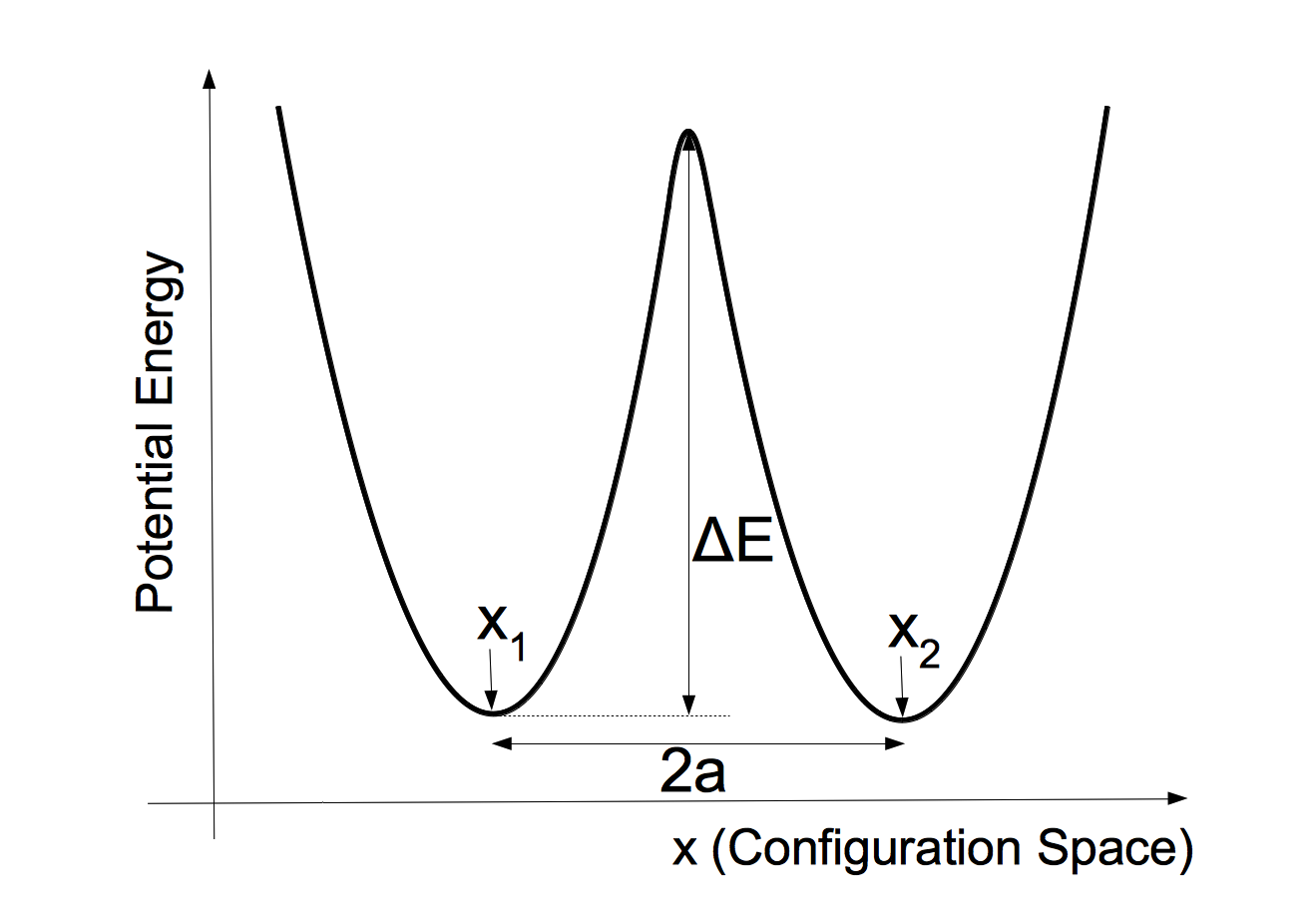}
  \caption{Two level systems where structural relaxation is achieved
   through a jump from one minimum to the other one overcoming an 
   energy barrier $\Delta E$ (see text for details).}
  \label{fig:shovingmodel}
\end{figure}

\begin{figure}[h]
\centering
  \includegraphics[width=\linewidth]{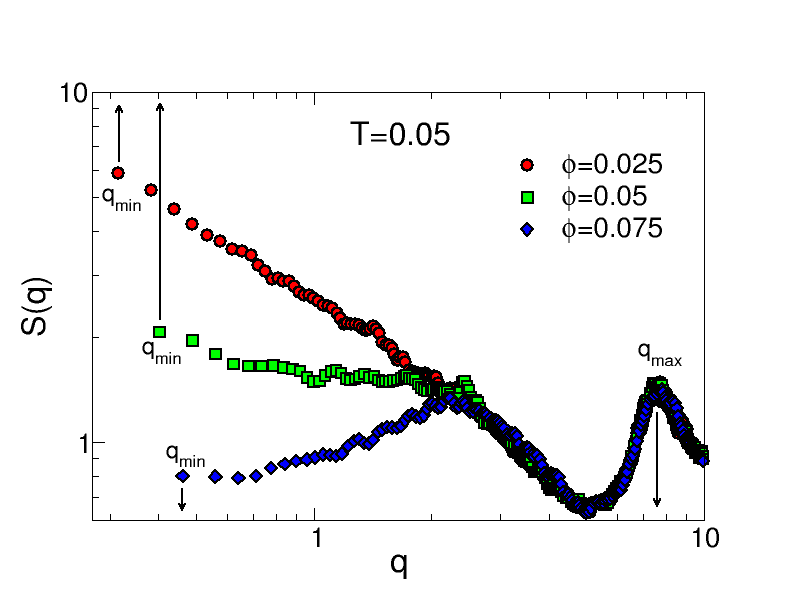}
  \caption{Static structure factor at lowest temperatures
  investigated for all volume fractions. $q_{min}$ and $q_{max}$ are
  are also pointed out where $q_{min}$ is the minimum
  wave vector allowed by finite size of simulation box and $q_{max}$
  is the wave vector corresponding to the maximum of $S(q)$ at
  length scales comparable to particles diameter. }
  \label{fig:Sq}
\end{figure}

\begin{figure}[h]
\centering
  \includegraphics[width=\linewidth]{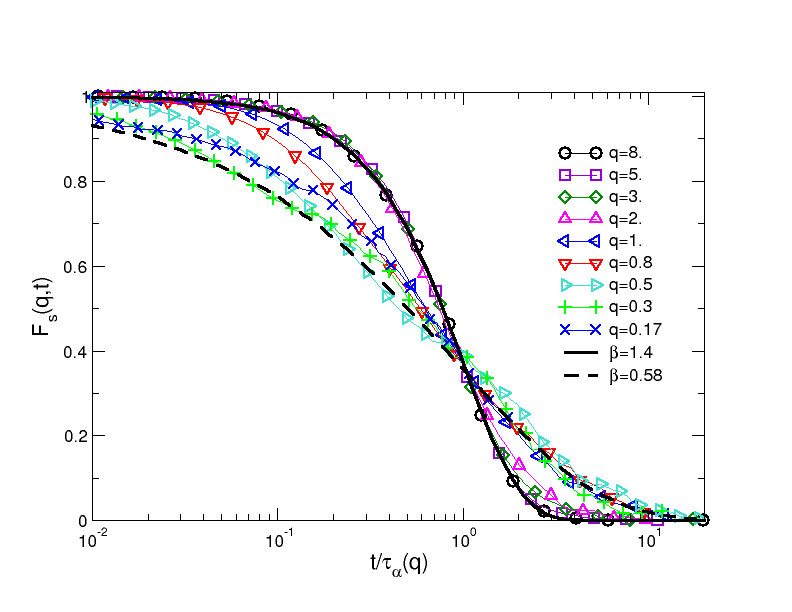}
  \caption{Self part of intermediate scattering function for different $q$ from $q_{min}$ to values around 
  $q_{max}$. Solid and dashed black lines are fits to a stretched exponential with $\beta=0.58$ (dashed) and
  $\beta=1.4$ (solid) for minimum and maximum wave vectors considered here respectively. }
  \label{fig:Fqs}
\end{figure}

The paper is organized as follows. In Sec. \ref{sec:HWgeneq} we recall the basics to derive 
the  universal scaling form subsequently discussed,
in Sec. \ref{sec:methods} we briefly summarize the main features of the gel model 
as studied in Ref. \cite{edgkob_epl05,edgkob_prl07,edgkob_sm10} 
by molecular dynamics and of numerical simulations.
In Sec. \ref{sec:resdisc} we provide a simple protocol to test the universal scaling\cite{OurNatPhys} for the gel system studied here. Finally in Sec. \ref{sec:conclusions} we draw the conclusions of our analysis.

\section{Universal Scaling Form}
\label{sec:HWgeneq}
On approaching the glass transition particles are longer and longer trapped into
the cage formed by their neighbors.
Caging phenomenon shows up as a plateau-like regime at short times in the MSD. 
The amplitude of the rattling motion $\langle u^2\rangle^{1/2}$ during this caging regime,
that occurs on very short time scales (e.g. picoseconds in molecular liquids),  is the so-called Debye-Waller (DW) factor \cite{StarrEtAl02,Ngai04} which is directly related to the short-time elastic properties of the system \cite{Dyre06}. 

The DW factor is an experimentally accessible quantity \cite{Ngai00} that
can be also measured by using the incoherent intermediate scattering function (ISF),
evaluating at the short times the height of the plateau which signals the cage effects (see 
Ref. \cite{JCPwoly2009}).
We note that as shown in \cite{JCPwoly2009} DW factor extracted from MSD $\langle u^2 \rangle $ and one defined from ISF are equivalent.

In spite of the fact that the DW factor is related to fast motion of particles occurring on time scales much shorter than the ones typical of structural relaxation, 
 many studies evidence in glass forming liquids a possible relation between slow and
 fast degrees of freedom \cite{TobolskyEtAl43,Harrowell06,VogGlotz04,ScopignoEtAl03,buchenau,Paciaroni05,Magazu04,Ngai04,Ngai00,Angell95,SokolPRL,NoviSoko04,KobFraPRLDW01}.

In order to express the correlation between DW factor and structural relaxation time in a functional form, a classical argument estimating the height of the barrier between two potential energy minima from the curvature around the minima can be used.
For glassy systems Hall and Wolynes \cite{HallWoly87} applied this argument
in their density functional theory where atomic motion is restricted to cells, picturing the GT as a freezing in an aperiodic crystal structure.
In this approach system relaxation towards equilibrium can be thought as a series of activated
jumps over energy barriers in its potential energy landscape \cite{Dyre04}.
Following Ref. \cite{Dyre04} we now give a derivation of an equation relating
DW factor and structural relaxation, which is useful in the context of this paper.


For the sake of simplicity we restrict to the one-dimensional case where two minima
are separated by a distance $2a$ (see Fig. \ref{fig:shovingmodel}).
Referring to Fig. \ref{fig:shovingmodel},
we expand the potential  $U(x)$ around the minimum on the left, whose position is labeled by $x_1$:
\begin{equation}
U(x) = U_0 + \frac{\Lambda}{2} (x-x_1)^2 
\label{eq:Uexp}
\end{equation}
Since system relaxation requires getting over the energy barrier $\Delta E$, if $\tau_\alpha$
is the system relaxation time and $\tau_0$ the microscopic time: 
\begin{equation}
\tau_\alpha = \tau_0 \exp \left ( \frac{\Delta E}{k_B T} \right )
\label{eq:activrel}
\end{equation}
From Eq. (\ref{eq:Uexp}) we can express $\Delta E$ as:
\begin{equation}
\Delta E = \frac{\Lambda}{2} a^2
\label{eq:DeltaE} 
\end{equation}
and from equipartition theorem:
\begin{equation}
k_B\, T = \Lambda \langle u^2 \rangle
\label{eq:equip}
\end{equation}
where $\langle u^2 \rangle = \langle x^2 \rangle$.
Inserting Eqs. (\ref{eq:equip}) and (\ref{eq:DeltaE}) into Eq. (\ref{eq:activrel})
one obtains finally:
\begin{equation}
\tau_\alpha = \tau_0 \exp \left ( \frac{a^2}{2\, \langle u^2 \rangle}\right )
 \label{eq:origHW}
\end{equation}
It is important to note that Eq. (\ref{eq:origHW}) is expected to fail
if  the amplitude of rattling motion $\langle u^2 \rangle$ becomes comparable
to $a^2$.

A natural generalization of Eq.(\ref{eq:origHW}) can be achieved adopting
a proper distribution $p(a^2)$ of the squared displacement $a^2$
needed to overcome energy barriers, i.e. in our present study to break bonds.
We note that the squared displacement $a^2$ is the cumulative displacement of the particles that move \cite{HallWoly87}, hence according to Central Limit Theorem a suitable choice for $p(a^2)$ is a truncated 
gaussian form, i.e.
\begin{equation}
p(a^2) = \left\{ \begin{array}{ll}
A \exp\left [-\frac{(a^2 - \overline{a^2})^2}{2\sigma^2_{a^2}}\right ]
 & \textrm{if $a>a_{min}$}\\
0& \textrm{otherwise}
\end{array} \right.
\label{eq:gaussDistro}
\end{equation}
where $A$ is a normalization factor and $a^2_{min}$ is the minimum displacement to reach the transition state. Averaging the Eq.
(\ref{eq:origHW}) over the distribution given by Eq. (\ref{eq:gaussDistro}),  the following generalized HW equation is obtained:
\begin{equation}
\tau_\alpha =  \tau_0 \;  \exp\left ( \frac{\overline{a^2}}{2\langle u^2\rangle } + \frac{ \sigma^2_{a^2}}{8\langle u^2\rangle ^2 } \right )
\label{eq:extHW}
\end{equation}

The gaussian form for $p(a^2)$ is supported also by other considerations.
For example if we substitute back $k_B T$ into Eq. (\ref{eq:extHW}) using
Eq. (\ref{eq:equip}) we end up with the following equation:
\begin{equation}
\tau_\alpha = \tau_0\exp\left [\frac{\Lambda\, \overline{a^2}}{2 k_B T} + \frac{\Lambda^2\, \sigma^2_{a^2}}
{8 (k_B T)^2} \right ]
\label{eq:harm}
\end{equation}
Experimental data for both supercooled liquids \cite{Bassler87} and polymers \cite{FerryEtal53} together with theoretical approaches \cite{EastArrow} support  gaussian
form for $p(a^2)$ in Eq. (\ref{eq:harm}).
Furthermore using Eq. (\ref{eq:DeltaE}) into Eq. (\ref{eq:gaussDistro}) 
to eliminate $a^2$ a gaussian distribution for energy barriers is attained 
in accordance with other studies \cite{MonthBouch96}.

These ideas have been originally developed for the glassy dynamics of dense systems, where the caging occurs due to the high density. In the following we would like to try and apply them to the glassy dynamics of colloidal gels, where some hints of caging phenomena appear, although densities can be very low. The caging in these systems rather originates from the formation of persistent bonds \cite{DeMicheleSticky06}, which eventually lead to an interconnected network structure.
\section{Methods}
\label{sec:methods}
\subsection{Model}

We refer to the studies carried on in Refs.\cite{edgkob_epl05,edgkob_prl07,edgkob_sm10}:
the colloidal gel model considers identical particles of diameter $\sigma$ 
interacting via a phenomenological potential $V_{eff}$, designed to 
account for the presence of directional interactions.  
In gelling colloidal suspensions there are in fact several possible sources of 
anisotropic effective interactions, since the particle surface may not be smooth or 
the building blocks of the gel are not the primary particles but larger aggregates of irregular shape \cite{exp2}. 
Confocal microscopy images obtained in recent experiments 
\cite{dinsmore_prl06,solomon,royall} confirm this scenario: the distribution of the particle coordination number $n$ in very diluted gel networks is strongly peaked around $n \simeq 2,3$.
In the chosen model, therefore, the interaction potential is given as the 
sum of three different contributions, $V_{eff}= V_{LJ} + V_{d} + V_{3}$,
where $V_{LJ}$ is a Lennard-Jones type of potential producing a
narrow attractive well, and $V_{d} + V_{3}$ contains directionality and rigidity 
of inter-particle bonding \cite{edgkob_sm10}.
Here we consider the same choice of parameters as in 
\cite{edgkob_epl05,edgkob_prl07,edgkob_sm10} 
and the range of volume fractions as investigated in \cite{edgkob_sm10}. 

As reported in the previous studies, in this model at low temperatures the system aggregates into an open persistent network of chains connected by a few bridging points ({\it nodes}). 
This takes place via a random percolation mechanism, but once a 
percolating structure is formed,
it rapidly evolves towards a persistent, fully connected open network.
The formation of the persistent network produces the coexistence,
in the gel, of very different relaxation processes at different length
scales: the relaxation at high wave vectors is due to the fast cooperative
motion of pieces of the gel structure (e.g. the chains connecting two nodes), whereas at low wave vectors the overall rearrangements of the heterogeneous gel make the system relax via a stretched exponential decay of the time correlators. The coexistence of such diverse relaxation mechanisms is characterized by a typical crossover length which is of the order of the network mesh size. The slow glassy dynamics at low wave vectors results to be directly connected to the presence of cooperative processes which can be recognized, for example, 
in the rearrangements of the network nodes along the complex structure 
of the network itself \cite{edgkob_sm10}.
This scenario is in agreement with the results of other recent studies on model colloidal gels \cite{miller,kob_sastry_09}.

\subsection{Simulations details}
We have used a MD code where the potential $V_{eff}$ has been 
implemented via a suitable combination of the algorithms RATTLE 
and SHAKE~\cite{md}.
The unit of time is $\sqrt{m \sigma^2/\epsilon}$, with $m$ the mass of a
particle and the data reported here refer to a time step of $0.002$.
The data refers to micro-canonical simulations performed with 8000 particles in cubic boxes of size $L=37.64, 43.09, 55.10$ in unit 
of $\sigma$, corresponding respectively to particle densities of 
$\rho = 0.15$, $0.1$, and $0.05$, i.e. to  
approximately volume fractions $\phi \simeq 0.075$, $0.05$, and $0.025$.
In the simulations 5 to 8 independent samples have been equilibrated 
starting from initial high temperature random configurations 
by replacing particle velocities with values extracted from a Maxwell-Boltzmann distribution every $\Delta$ time steps (where $\Delta$ varied with 
temperature from $10$ to $10^3$ MD steps).
After equilibration the energy is constant, showing no significant drift over the simulation time window, and different one- and two- time autocorrelation 
functions display the equilibrium behavior, i.e. do not show any sign of aging.
The data production starts from these equilibrated samples: 
the equilibration time grows with the relaxation time in the system and at 
the lowest temperatures equilibration required up to $2\cdot10^7$ MD steps.

\section{Results and Discussion}
\label{sec:resdisc}
\subsection{Relaxation and Transport Properties}
\label{sec:relaxtransprop}
We use the static structure factor $S(q)$, defined as follows:
\begin{equation}
S(q) = \frac{1}{N}\sum_{i,j} \langle e^{i {\bf q} ( {\bf r}_i - {\bf r}_j} ) \rangle
\label{eq:Sq}
\end{equation}
to quantify the extent of spatial correlation in the system and obtain informations on the gel structure.
In Fig. \ref{fig:Sq} (from the data of Ref.\cite{edgkob_sm10}) 
$S(q)$ of the gel network (i.e. at the lowest temperature considered)
displays a peak around $q_{max}\approx 8 $ corresponding roughly to the particle 
diameter. This peak basically arises from excluded volume interactions between 
particles, i.e. it approximately corresponds to the first peak of radial 
distribution function. In glassy systems the slow relaxation
arises first, and has its strongest signature, at these wave vectors. 
It is clear from the figure that in the gel significant spatial correlations
are present also at smaller wave vectors.
In Ref.\cite{edgkob_sm10} $S(q)$ has been compared to the static 
structure factor of a polymer chain solution\cite{rubinstein}: length scales matching smaller 
wave vectors $2.0 < q < 7.0$ can interpreted as an {\it intra-molecular} 
regime for spatial correlations of the aggregates (i.e. chains). 
Mesoscopic and macroscopic length scales $q \leq 2.0$ can be instead thought 
of as corresponding to {\it inter-molecular} regime, due to the long-range
interactions induced by the formation of the persistent gel network.
Correlations in the particle motion over different length scales 
can be effectively quantified in terms of ISF:
\begin{equation}
F_s (q,t) = \frac{1}{N} \sum_{j}^N \langle e^{i {\bf q} [{\bf r}_i(t)-{\bf r}_j(0)]} \rangle
\end{equation}
The analysis of its behavior  \cite{edgkob_epl05,edgkob_prl07,edgkob_sm10} 
indicates that in the gel the slowest modes correspond to the 
{\it inter-molecular} regime of wave vectors. 
In Fig. \ref{fig:Fqs} 
$F_{s}(q,t)$ is plotted as a function of the time, rescaled by the relaxation 
time $\tau_{\alpha}(q)$, at the lowest temperature $T=0.05$ for different wave 
vectors. $\tau_{\alpha}(q)$ has been calculated from $F_{s}(q,\tau_{\alpha}(q))=1/e$.
The figure well shows that 
the stretched exponential decay $\exp \{-[t/\tau_{\alpha}(q) ]^{\beta}\}$ with
$\beta < 1.$
typical of glassy dynamics, arises only at low wave vectors ($q \le 1.0$).
In other words, the slow structural relaxation modes 
of the gel structure can be detected only at low $q$.
In contrast, at higher wave vectors, the time decay of correlations 
is faster than exponential ($\beta \simeq 1.4$): 
in Refs.\cite{edgkob_prl07,edgkob_jnnfm08}
these processes have been carefully analyzed and it has been shown that they 
are due to fast coherent motion of pieces of the gel network 
(i.e. the chains between two bridging point or {\it nodes}).
 

The overall scenario of relaxation modes in the gel is therefore rather 
different from the one discussed in 
\cite{OurNatPhys,JCPwoly2009} for glassy systems at high densities. 
It is interesting to notice that here the particle bonding is the
basic ingredient of the structural arrest, since the 
persistence of the gel network certainly relies upon the long living bonds.
On the other hand, the analysis of the relaxation modes well elucidate that
formation of single bonds cannot be responsible, on its own, for the 
cooperative glassy dynamics which instead arises from the long-range 
correlations between them induced by the network \cite{edgkob_sm10}.

Within this picture, the question of the existence and nature 
of a direct correlation
between the single particle average localization and the structural arrest,
as discussed in Sec.\ref{sec:univscal}, becomes particularly intriguing 
and is the main focus of this work.  

To this aim, we have calculated the relaxation time $\tau_{\alpha}$ associated 
to the structural relaxation of the gel as 
$\tau_{\alpha}=\tau_{\alpha}(q_{min})$, where $q_{min}$ is the smallest wave 
vector compatible with periodic boundary conditions in our simulations boxes, 
i.e. it corresponds to length scales of the order of the box size. 


For evaluating the rattling amplitude of the particle motion corresponding to caging, 
we consider the MSD:
\begin{equation}
\langle r^2(t)\rangle =  \frac{1}{N} \sum_i^N \langle \| {\bf r}_i(t) - {\bf r}_i(0) \|
\rangle.
\label{eq:MSD}
\end{equation}
Figure \ref{fig:MSD} shows MSD divided by time for all temperatures at volume fraction 
$\phi=0.05$ (from Ref.\cite{edgkob_epl05}). Since the system is very diluted, localization phenomena can be very weak and this type of plot helps to better recognize different regimes.
At very short times a ballistic regime is 
found where MSD increases according to 
$\langle r^2(t) \rangle\simeq (3 k_b T / m) t^2$ (i.e.$\langle r^2(t) \rangle/t\propto t$). 
Formation of bonds with other particles slows down the displacement and 
$\langle r^2(t) \rangle/t$ shows an inflection point.
At high temperatures, bonds break within a time interval much smaller than 
$\tau_\alpha$ and particle starts diffusing, 
i.e. $\langle r^2(t)\rangle /t$ eventually reaches a plateau. At $T < 0.1$ 
bond lifetime sets instead the longest relaxation time-scale in the 
system and in this regime the formation of the gel network starts, 
with the MSD becoming increasingly sub-diffusive over times much longer than
the localization process related to the rattling of the particle within the
bonding length scale. Therefore we evaluate the caging from this first 
localization process as explained in the following. 

\subsection{Scaling between relaxation and caged dynamics}
\label{sec:univscal}
\begin{figure}[h]
\centering
  \includegraphics[width=\linewidth]{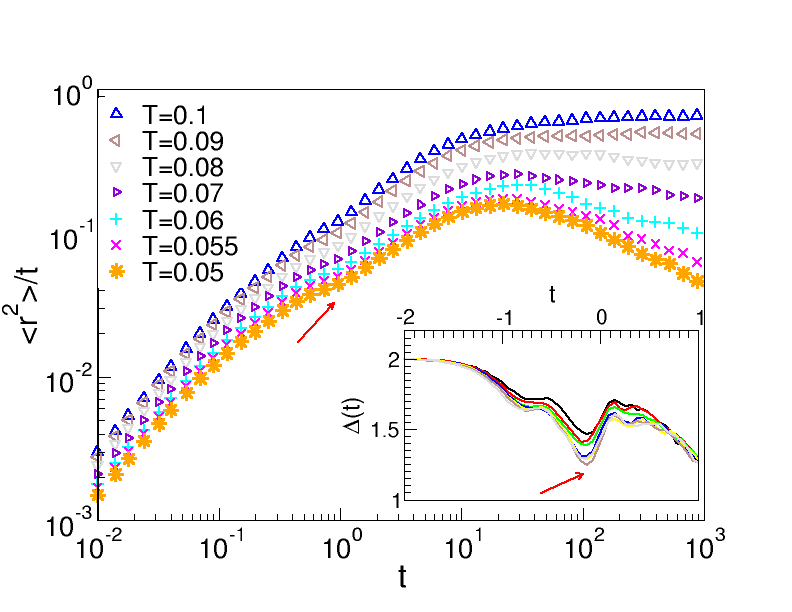}
  \caption{$\langle r^2(t)\rangle/t$ as a function of time at $\phi=0.05$. Inset: 
  Logarithmic derivative showing the minimum corresponding to caging. 
The arrows indicate the regime chosen for the evaluation of DW.}
  \label{fig:MSD}
\end{figure}

Following the discussion 
in Ref.\cite{JCPwoly2009}, we evaluate the DW 
factor in our gel system in order to provide a suitable 
characteristic length scale for the particle temporarily 
trapped into the cage due to bonds formation.
DW factor can be defined picking a suitable value of MSD within a 
time window that begins just after ballistic regime and that ends 
just before structural relaxation sets in.
First we have to identify such time window and to do that
we consider the slope $\Delta(t)$ of MSD in a log-log plot, i.e.:
\begin{equation}
\Delta(t) = \frac{\partial \log \langle r^2\rangle}{\partial \log t}
\label{eq:MSDslope}
\end{equation}
Representative plots of $\Delta(t)$ for our gel system can be found
in the inset of Fig. \ref{fig:MSD}.
The short-time ballistic regime corresponds to $\Delta(t) \approx 2$
while the long time diffusive regime corresponds to $\Delta(t) \approx 1$.
Between these two regimes an intermediate regime is present where
caging of particles due to bonds gives rise to a clear 
minimum of $\Delta(t)$ (see Fig. \ref{fig:MSD}). 
We thus define the DW factor $\langle u^2\rangle$ as follows:
\begin{equation}
\langle u^2\rangle = \langle r^2(t=t^*) \rangle
\label{eq:DWdefinition}
\end{equation}
where $t^*$ is the time corresponding to the minimum of $\Delta(t)$ within
this intermediate regime.
 
\begin{figure}[h]
\centering
  \includegraphics[width=\linewidth]{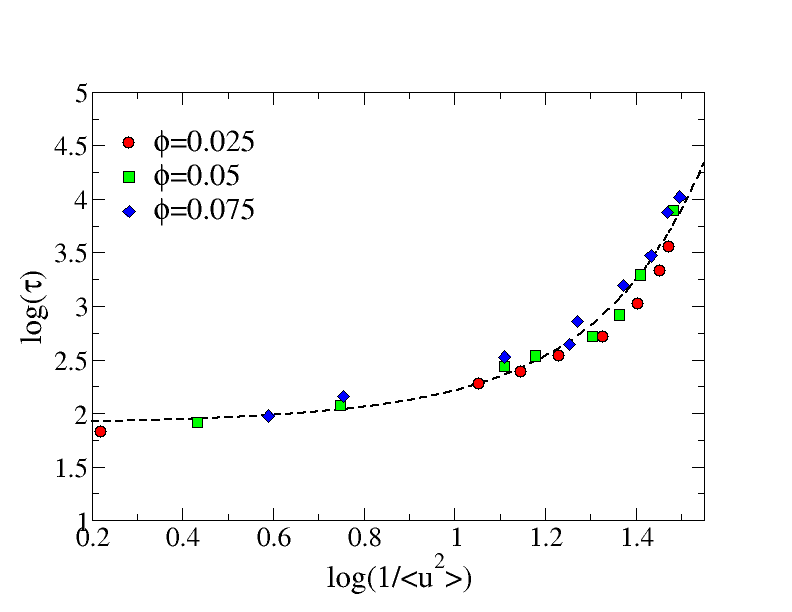}
  \caption{Scaling of gel data for all densities investigated, i.e.
  $\phi=0.025,0.05, 0.075$. Dashed line is a fit of all data to the function
  $\log \tau_\alpha = \alpha + \beta \langle u^2 \rangle^{-1} + 
  \gamma \langle u^2 \rangle^{-2}$  with $\alpha =1.893$, $\beta= 0.0177$ 
  and $\gamma=0.00144$. }
  \label{fig:gelscaling}
\end{figure}

\begin{figure}[h]
\centering  
\includegraphics[width=\linewidth]{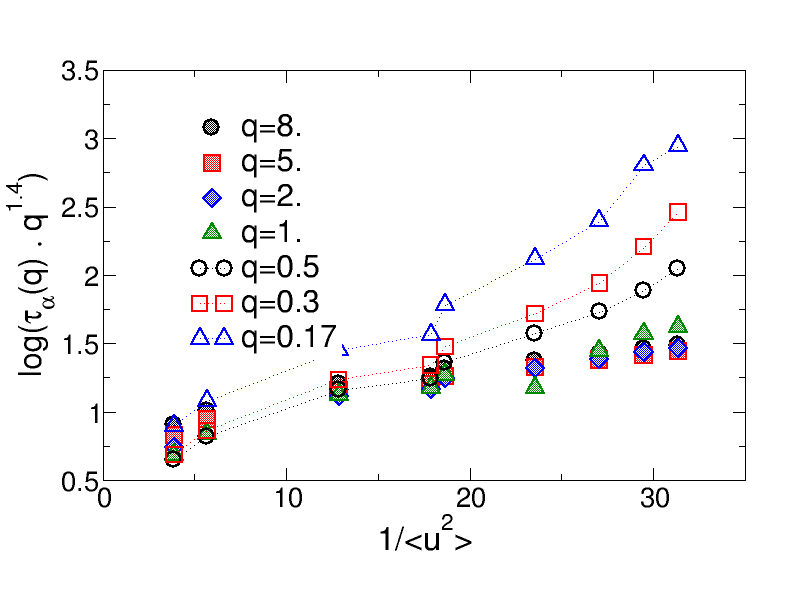}  
\caption{Plot of $log(\tau_{\alpha}(q)\cdot q^{1.4})$ versus $1/u^{2}$ for different 
$q$ at $\phi=0.075$. Upon increasing the wave vector from $q_{min}$, 
the data strongly depart from the scaling form of Eq.\ref{eq:extHW}. 
}  
\label{fig:gelscaling_q}
\end{figure}

We are now in a position to establish a correlation between 
structural relaxation and caged dynamics. In Fig. \ref{fig:gelscaling}
$\log(\tau_\alpha)$ is plotted against the inverse of 
DW factor $1/\langle u^2 \rangle $ for the three volume fractions investigated.
The figure clearly shows that
$\tau_\alpha$  is strongly correlated to the DW factor. The correlation has 
also a specific form (see the fitting curve in the figure) well agreeing with the
prediction of Eq. (\ref{eq:extHW}).
This is far from being obvious because here 
the structural relaxation is related to length scales of the order of 
simulation box (i.e. $2 \pi / q_{min}$) whereas 
the DW factor corresponds to caging phenomena 
occurring on much smaller length scales 
of the order of first neighbor distance, i.e. $q \approx 2 \pi / \sigma \approx q_{max}$ (see Fig. \ref{fig:Sq}),
where the relaxation is rather dominated by other mechanisms (see Fig.\ref{fig:Fqs}). 
It is also remarkable that the all data for different volume fractions collapse onto the same scaling curve, because the structure, as described by spatial correlations in Fig.\ref{fig:Sq}, 
changes significantly with $\phi$ at small $q$ ($S(q_{min})$ increases in fact by almost an order of magnitude from $\phi=0.075$ to $\phi=0.025$), whereas it is not affected by changes in $\phi$ at large wave vectors $q$.

The correlation between DW factor and structural 
relaxation time $\tau_\alpha$ points to a strong correlation between the 
long-range correlations established by the network and the localization  
within the structure at the level of the single particle.  

To better elucidate the nature of the scaling found, we have also investigated the existence of correlations at different wave vectors. In the gel network,  $\tau_{\alpha}(q)$ displays a complex 
dependence on $q$ as discussed in Refs.\cite{edgkob_prl07,edgkob_sm10}. In particular,  $\tau_{\alpha}(q) \propto q^{-1.4}$ at large $q$, corresponding to the regime where relaxation is dominated 
by fast collective motion of pieces of the structures (i.e. chains between two nodes). 
Therefore  in Fig.\ref{fig:gelscaling_q} we have used the same type of plot as in Fig.\ref{fig:gelscaling} where $\tau_\alpha(q)$ has been rescaled with $q^{1.4}$. The data refer 
to different $q$ at $\phi=0.075$. At the largest $q$ the data well collapse on top of each other 
and display a very different dependence on $1/\langle u^2 \rangle$. Upon decreasing $q$ the data depart from the $q^{1.4}$ scaling in $q$ and also approach the scaling form of Fig.\ref{fig:gelscaling}. 
This analysis further demonstrate that the scaling found specifically relates the average localization at the level of the single particle (as quantified by $\langle u^2 \rangle$) to the relaxation modes associated to the arising of  glassy, cooperative dynamics. 

\begin{figure}[h]
\centering
  \includegraphics[width=\linewidth]{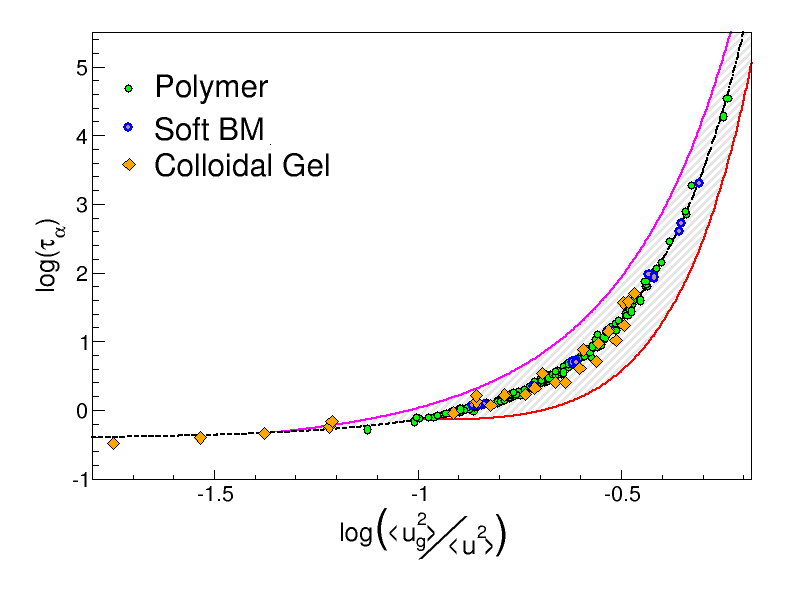}
  \caption{Scaling of the structural relaxation time vs the reduced
DW factor of polymers \cite{OurNatPhys}, soft binary mixtures (BM) \cite{JCPwoly2009} and colloidal gel (present work).
For the colloidal gel $\langle u_g^2 \rangle^{1/2} = 0.103$ and data have been shifted vertically by $-2.33$ in log scale. Dashed line is the universal curve defined in Eq. (\ref{eq:scaledparabola}).
Solid lines bound  the accuracy of  Eq.\ref{eq:extHW} obtained fitting data from polymer models studied
in \cite{OurNatPhys}  and correspond to the two definitions $\langle u^2\rangle \equiv  \langle r^2(t=0.6)\rangle$ (magenta) and $\langle u^2\rangle \equiv  \langle r^2(t=1.4)\rangle$ (red) (see \cite{JCPwoly2009} 
for more details). }
 \label{fig:univscal_gel}
\end{figure}



\subsubsection{Comparison with other MD Studies}

In Ref. \cite{JCPwoly2009} it has been shown that for several model glassformers and experimental systems
plotting $\log(\tau_\alpha)$ versus $\langle u_g^2 \rangle / \langle u^2\rangle$ (where 
$\langle u_g^2 \rangle$ is the DW factor at the GT)
all data scale onto the same master curve, i.e.:
 \begin{equation}
\log \tau_\alpha = \alpha + \tilde{\beta} \;  \frac{\langle u^2_g \rangle}{\langle u^2 \rangle} + \tilde{\gamma} \left (\frac{\langle u^2_g \rangle}{\langle u^2\rangle} \right )^2 
\label{eq:scaledparabola}
\end{equation}
where:  
\begin{eqnarray}
\alpha &=&-0.424(1) \\
\tilde{\beta} &=& \frac{\overline{a^2}}{2 \ln10 \langle u^2_g \rangle}= 1.62(6) \label{betatilde}  \\
\tilde{\gamma} &=&\frac{\sigma^2_{a^2}}{8 \ln10 \langle u^2_g \rangle^2} = 12.3(1) \label{gammatilde}
\end{eqnarray}


The scaling form obtained in Fig. \ref{fig:gelscaling} for the colloidal gel can in fact be superimposed on the universal curve of Eq. (\ref{eq:scaledparabola}) with a suitable vertical shift ($\alpha'=\alpha-2.33$) and upon using on $x$-axis the scaled variable 
$\langle u^2 \rangle/\langle u_g^2\rangle$ with $\langle u_g^2 \rangle = 0.104$
in order to rule out any trivial dependence on time and length scales.
Using such vertical shift and such value for $\langle u_g^2 \rangle$ to adjust gel data
we can compare them  to the results obtained from soft binary mixtures \cite{JCPwoly2009} and polymer systems \cite{OurNatPhys} as shown in Fig.\ref{fig:univscal_gel}.
It is clear from this figure that within the accuracy (marked by solid lines) the scaling procedure works well also for the colloidal gel model considered in the present paper.

\section{Conclusions}
\label{sec:conclusions}
We have investigated possible correlations between the localization at the level of the single particle and structural relaxation in a model colloidal gel at very low volume fractions, with directional effective 
interactions and local rigidity. We have found that strong correlations are present over different 
length scales. In this type of systems, the localization of particles due to persistent bonding is of course 
the initiator of the process that leads to gelation, but cannot be responsible, on its own, for structural arrest, which arises thanks to the formation, eventually, of an interconnected network structure.   
Remarkably, we have found that relaxation modes at the lowest wave vectors, i.e. over length scale much larger than the bond localization length, strongly correlate to the localization of single particles within the typical bonding length.  We think that these findings indicate a type of {\it feedback} mechanism between dynamical processes at different length-scales: particle bonding leads to the network formation and long range correlations induced by the presence of the network actually transforms the particle bonding into a glassy caging, coupling 
eventually the particle localization to the glassy structural arrest. 
Moreover, we have shown that the gel data display the same scaling form found for a large class of dense glassy systems in experiments and simulations (molecular glasses, polymers, etc. ): this result  
strongly suggests that the scaling captures the essential, basic ingredients in the physics of glassy structural arrest and it is an extremely powerful tool for devising its possible universal features. 

\section{Acknowledgments}
CDM acknowledges support from ERC (226207-PATCHYCOLLOIDS).





\footnotesize{
\bibliography{univscalgel} 

\providecommand*{\mcitethebibliography}{\thebibliography}
\csname @ifundefined\endcsname{endmcitethebibliography}
{\let\endmcitethebibliography\endthebibliography}{}
\begin{mcitethebibliography}{63}
\providecommand*{\natexlab}[1]{#1}
\providecommand*{\mciteSetBstSublistMode}[1]{}
\providecommand*{\mciteSetBstMaxWidthForm}[2]{}
\providecommand*{\mciteBstWouldAddEndPuncttrue}
  {\def\EndOfBibitem{\unskip.}}
\providecommand*{\mciteBstWouldAddEndPunctfalse}
  {\let\EndOfBibitem\relax}
\providecommand*{\mciteSetBstMidEndSepPunct}[3]{}
\providecommand*{\mciteSetBstSublistLabelBeginEnd}[3]{}
\providecommand*{\EndOfBibitem}{}
\mciteSetBstSublistMode{f}
\mciteSetBstMaxWidthForm{subitem}
{(\emph{\alph{mcitesubitemcount}})}
\mciteSetBstSublistLabelBeginEnd{\mcitemaxwidthsubitemform\space}
{\relax}{\relax}

\bibitem[Angell(1995)]{Angell95}
C.~A. Angell, \emph{Science}, 1995, \textbf{267}, 1924--1935\relax
\mciteBstWouldAddEndPuncttrue
\mciteSetBstMidEndSepPunct{\mcitedefaultmidpunct}
{\mcitedefaultendpunct}{\mcitedefaultseppunct}\relax
\EndOfBibitem
\bibitem[Debenedetti and Stillinger(2001)]{DebeStilli2001}
P.~G. Debenedetti and F.~H. Stillinger, \emph{Nature}, 2001, \textbf{410},
  259--267\relax
\mciteBstWouldAddEndPuncttrue
\mciteSetBstMidEndSepPunct{\mcitedefaultmidpunct}
{\mcitedefaultendpunct}{\mcitedefaultseppunct}\relax
\EndOfBibitem
\bibitem[Frenkel(2002)]{frenkel}
D.~Frenkel, \emph{Science}, 2002, \textbf{296}, 65\relax
\mciteBstWouldAddEndPuncttrue
\mciteSetBstMidEndSepPunct{\mcitedefaultmidpunct}
{\mcitedefaultendpunct}{\mcitedefaultseppunct}\relax
\EndOfBibitem
\bibitem[Pham \emph{et~al.}(2002)Pham, Puertas, Bergenholz, S.~U.~Egelhaaf,
  Pusey, Schofield, Cates, Fuchs, and Poon]{pham}
K.~N. Pham, A.~M. Puertas, J.~Bergenholz, A.~M. S.~U.~Egelhaaf, P.~Pusey, A.~B.
  Schofield, M.~Cates, M.~Fuchs and W.~Poon, \emph{Science}, 2002,
  \textbf{296}, 5565\relax
\mciteBstWouldAddEndPuncttrue
\mciteSetBstMidEndSepPunct{\mcitedefaultmidpunct}
{\mcitedefaultendpunct}{\mcitedefaultseppunct}\relax
\EndOfBibitem
\bibitem[Anyfantakis \emph{et~al.}(2009)Anyfantakis, Bourlinos, Vlassopoulos,
  Fytas, Giannelis, and Kumar]{vlassopoulos}
M.~Anyfantakis, A.~Bourlinos, D.~Vlassopoulos, G.~Fytas, E.~Giannelis and S.~K.
  Kumar, \emph{Soft Matter}, 2009, \textbf{5}, 4256\relax
\mciteBstWouldAddEndPuncttrue
\mciteSetBstMidEndSepPunct{\mcitedefaultmidpunct}
{\mcitedefaultendpunct}{\mcitedefaultseppunct}\relax
\EndOfBibitem
\bibitem[Ramos and Cipelletti(2005)]{luca_jpcm05}
L.~Ramos and L.~Cipelletti, \emph{Journal of Physics: Condensed Matter}, 2005,
  \textbf{17}, R253\relax
\mciteBstWouldAddEndPuncttrue
\mciteSetBstMidEndSepPunct{\mcitedefaultmidpunct}
{\mcitedefaultendpunct}{\mcitedefaultseppunct}\relax
\EndOfBibitem
\bibitem[{S. M.}~Fielding and Sollich(2009)]{SollichEtAl09}
M.~C. {S. M.}~Fielding and P.~Sollich, \emph{Soft Matter}, 2009, \textbf{5},
  2378\relax
\mciteBstWouldAddEndPuncttrue
\mciteSetBstMidEndSepPunct{\mcitedefaultmidpunct}
{\mcitedefaultendpunct}{\mcitedefaultseppunct}\relax
\EndOfBibitem
\bibitem[Zaccarelli(2007)]{EZJPCM07}
E.~Zaccarelli, \emph{J. Phys.: Condens. Matter}, 2007, \textbf{19},
  323101\relax
\mciteBstWouldAddEndPuncttrue
\mciteSetBstMidEndSepPunct{\mcitedefaultmidpunct}
{\mcitedefaultendpunct}{\mcitedefaultseppunct}\relax
\EndOfBibitem
\bibitem[Likos(2001)]{LikosPhysRep01}
C.~N. Likos, \emph{Physics Reports}, 2001, \textbf{348}, 267 -- 439\relax
\mciteBstWouldAddEndPuncttrue
\mciteSetBstMidEndSepPunct{\mcitedefaultmidpunct}
{\mcitedefaultendpunct}{\mcitedefaultseppunct}\relax
\EndOfBibitem
\bibitem[Tobolsky \emph{et~al.}(1943)Tobolsky, Powell, and
  Eyring]{TobolskyEtAl43}
A.~Tobolsky, R.~E. Powell and H.~Eyring, Frontiers in Chemistry, New York,
  1943, pp. 125--190\relax
\mciteBstWouldAddEndPuncttrue
\mciteSetBstMidEndSepPunct{\mcitedefaultmidpunct}
{\mcitedefaultendpunct}{\mcitedefaultseppunct}\relax
\EndOfBibitem
\bibitem[Angell(1968)]{Angell68}
C.~A. Angell, \emph{J. Am. Chem. Soc.}, 1968, \textbf{86}, 117--124\relax
\mciteBstWouldAddEndPuncttrue
\mciteSetBstMidEndSepPunct{\mcitedefaultmidpunct}
{\mcitedefaultendpunct}{\mcitedefaultseppunct}\relax
\EndOfBibitem
\bibitem[S.V.Nemilov(1968)]{Nemilov68}
S.V.Nemilov, \emph{Russ. J. Phys. Chem.}, 1968, \textbf{42}, 726--729\relax
\mciteBstWouldAddEndPuncttrue
\mciteSetBstMidEndSepPunct{\mcitedefaultmidpunct}
{\mcitedefaultendpunct}{\mcitedefaultseppunct}\relax
\EndOfBibitem
\bibitem[Hall and Wolynes(1987)]{HallWoly87}
R.~W. Hall and P.~G. Wolynes, \emph{J. Chem. Phys.}, 1987, \textbf{86},
  2943--2948\relax
\mciteBstWouldAddEndPuncttrue
\mciteSetBstMidEndSepPunct{\mcitedefaultmidpunct}
{\mcitedefaultendpunct}{\mcitedefaultseppunct}\relax
\EndOfBibitem
\bibitem[Buchenau and Zorn(1992)]{BuchZorn92}
U.~Buchenau and R.~Zorn, \emph{Europhys. Lett.}, 1992, \textbf{18},
  523--528\relax
\mciteBstWouldAddEndPuncttrue
\mciteSetBstMidEndSepPunct{\mcitedefaultmidpunct}
{\mcitedefaultendpunct}{\mcitedefaultseppunct}\relax
\EndOfBibitem
\bibitem[Ngai(2000)]{Ngai00}
K.~L. Ngai, \emph{J. Non-Cryst. Solids}, 2000, \textbf{275}, 7--51\relax
\mciteBstWouldAddEndPuncttrue
\mciteSetBstMidEndSepPunct{\mcitedefaultmidpunct}
{\mcitedefaultendpunct}{\mcitedefaultseppunct}\relax
\EndOfBibitem
\bibitem[Starr \emph{et~al.}(2002)Starr, Sastry, Douglas, and
  Glotzer]{StarrEtAl02}
F.~Starr, S.~Sastry, J.~F. Douglas and S.~Glotzer, \emph{Phys. Rev. Lett.},
  2002, \textbf{89}, 125501\relax
\mciteBstWouldAddEndPuncttrue
\mciteSetBstMidEndSepPunct{\mcitedefaultmidpunct}
{\mcitedefaultendpunct}{\mcitedefaultseppunct}\relax
\EndOfBibitem
\bibitem[Novikov and Sokolov(2003)]{NoviSoko03}
V.~N. Novikov and A.~P. Sokolov, \emph{Phys.Rev.E}, 2003, \textbf{67},
  031507\relax
\mciteBstWouldAddEndPuncttrue
\mciteSetBstMidEndSepPunct{\mcitedefaultmidpunct}
{\mcitedefaultendpunct}{\mcitedefaultseppunct}\relax
\EndOfBibitem
\bibitem[Scopigno \emph{et~al.}(2003)Scopigno, Ruocco, Sette, and
  Monaco]{ScopignoEtAl03}
T.~Scopigno, G.~Ruocco, F.~Sette and G.~Monaco, \emph{Science}, 2003,
  \textbf{302}, 849--852\relax
\mciteBstWouldAddEndPuncttrue
\mciteSetBstMidEndSepPunct{\mcitedefaultmidpunct}
{\mcitedefaultendpunct}{\mcitedefaultseppunct}\relax
\EndOfBibitem
\bibitem[Novikov and Sokolov(2004)]{NoviSoko04}
V.~N. Novikov and A.~P. Sokolov, \emph{Nature}, 2004, \textbf{431},
  961--963\relax
\mciteBstWouldAddEndPuncttrue
\mciteSetBstMidEndSepPunct{\mcitedefaultmidpunct}
{\mcitedefaultendpunct}{\mcitedefaultseppunct}\relax
\EndOfBibitem
\bibitem[Ngai(2004)]{Ngai04}
K.~L. Ngai, \emph{Phil. Mag.}, 2004, \textbf{84}, 1341--1353\relax
\mciteBstWouldAddEndPuncttrue
\mciteSetBstMidEndSepPunct{\mcitedefaultmidpunct}
{\mcitedefaultendpunct}{\mcitedefaultseppunct}\relax
\EndOfBibitem
\bibitem[Dyre and Olsen(2004)]{Dyre04}
J.~C. Dyre and N.~B. Olsen, \emph{Phys. Rev. E}, 2004, \textbf{69},
  042501\relax
\mciteBstWouldAddEndPuncttrue
\mciteSetBstMidEndSepPunct{\mcitedefaultmidpunct}
{\mcitedefaultendpunct}{\mcitedefaultseppunct}\relax
\EndOfBibitem
\bibitem[Dyre(2006)]{Dyre06}
J.~C. Dyre, \emph{Rev. Mod. Phys.}, 2006, \textbf{78}, 953--972\relax
\mciteBstWouldAddEndPuncttrue
\mciteSetBstMidEndSepPunct{\mcitedefaultmidpunct}
{\mcitedefaultendpunct}{\mcitedefaultseppunct}\relax
\EndOfBibitem
\bibitem[Widmer-Cooper \emph{et~al.}(2008)Widmer-Cooper, Perry, Harrowell, and
  Reichman]{Harrowell_NP08}
A.~Widmer-Cooper, H.~Perry, P.~Harrowell and D.~R. Reichman, \emph{Nat. Phys.},
  2008, \textbf{4}, 711--715\relax
\mciteBstWouldAddEndPuncttrue
\mciteSetBstMidEndSepPunct{\mcitedefaultmidpunct}
{\mcitedefaultendpunct}{\mcitedefaultseppunct}\relax
\EndOfBibitem
\bibitem[Yannopoulos and Johari(2006)]{Johari06}
S.~N. Yannopoulos and G.~P. Johari, \emph{Nature}, 2006, \textbf{442},
  E7--E8\relax
\mciteBstWouldAddEndPuncttrue
\mciteSetBstMidEndSepPunct{\mcitedefaultmidpunct}
{\mcitedefaultendpunct}{\mcitedefaultseppunct}\relax
\EndOfBibitem
\bibitem[Larini \emph{et~al.}(2008)Larini, Ottochian, De{ }Michele, and
  Leporini]{OurNatPhys}
L.~Larini, A.~Ottochian, C.~De{ }Michele and D.~Leporini, \emph{Nat. Phys.},
  2008, \textbf{4}, 42--45\relax
\mciteBstWouldAddEndPuncttrue
\mciteSetBstMidEndSepPunct{\mcitedefaultmidpunct}
{\mcitedefaultendpunct}{\mcitedefaultseppunct}\relax
\EndOfBibitem
\bibitem[Ottochian \emph{et~al.}(2009)Ottochian, {De~Michele}, and
  Leporini]{OttochianEtAl09}
A.~Ottochian, C.~{De~Michele} and D.~Leporini, \emph{J. Chem. Phys.}, 2009,
  \textbf{131}, 224517\relax
\mciteBstWouldAddEndPuncttrue
\mciteSetBstMidEndSepPunct{\mcitedefaultmidpunct}
{\mcitedefaultendpunct}{\mcitedefaultseppunct}\relax
\EndOfBibitem
\bibitem[Ottochian and Leporini()]{OttochianLepoJNCS10}
A.~Ottochian and D.~Leporini, \emph{J. Non-Cryst. Solids}, DOI:
  10.1016/j.jnoncrysol.2010.05.094\relax
\mciteBstWouldAddEndPuncttrue
\mciteSetBstMidEndSepPunct{\mcitedefaultmidpunct}
{\mcitedefaultendpunct}{\mcitedefaultseppunct}\relax
\EndOfBibitem
\bibitem[Ottochian and Leporini(2010)]{OttochianLepoPhilMag10}
A.~Ottochian and D.~Leporini, \emph{Phil. Mag.}, in press\relax
\mciteBstWouldAddEndPuncttrue
\mciteSetBstMidEndSepPunct{\mcitedefaultmidpunct}
{\mcitedefaultendpunct}{\mcitedefaultseppunct}\relax
\EndOfBibitem
\bibitem[Ottochian \emph{et~al.}(2008)Ottochian, {De~Michele}, and
  Leporini]{OttochianLepoPhilMag08}
A.~Ottochian, C.~{De~Michele} and D.~Leporini, \emph{Phil. Mag.}, 2008,
  \textbf{88}, 4057--4062\relax
\mciteBstWouldAddEndPuncttrue
\mciteSetBstMidEndSepPunct{\mcitedefaultmidpunct}
{\mcitedefaultendpunct}{\mcitedefaultseppunct}\relax
\EndOfBibitem
\bibitem[Gado \emph{et~al.}(2004)Gado, Fierro, de~Arcangelis, and
  Coniglio]{edg_pre04}
E.~D. Gado, A.~Fierro, L.~de~Arcangelis and A.~Coniglio, \emph{Phys. Rev. E},
  2004, \textbf{69}, 051103\relax
\mciteBstWouldAddEndPuncttrue
\mciteSetBstMidEndSepPunct{\mcitedefaultmidpunct}
{\mcitedefaultendpunct}{\mcitedefaultseppunct}\relax
\EndOfBibitem
\bibitem[Cates \emph{et~al.}(2004)Cates, M.~Fuchs, Poon, and
  Puertas]{puertas04}
M.~E. Cates, K.~K. M.~Fuchs, W.~Poon and A.~Puertas, \emph{J. Phys.: Condensed
  Matter}, 2004, \textbf{16}, S4861\relax
\mciteBstWouldAddEndPuncttrue
\mciteSetBstMidEndSepPunct{\mcitedefaultmidpunct}
{\mcitedefaultendpunct}{\mcitedefaultseppunct}\relax
\EndOfBibitem
\bibitem[Puertas \emph{et~al.}(2005)Puertas, Fuchs, and Cates]{puertas05}
A.~Puertas, M.~Fuchs and M.~Cates, \emph{J. Phys. Chemistry B}, 2005,
  \textbf{109}, 6666\relax
\mciteBstWouldAddEndPuncttrue
\mciteSetBstMidEndSepPunct{\mcitedefaultmidpunct}
{\mcitedefaultendpunct}{\mcitedefaultseppunct}\relax
\EndOfBibitem
\bibitem[{De~Michele} \emph{et~al.}(2006){De~Michele}, Gabrielli, Tartaglia,
  and Sciortino]{DeMicheleSticky06}
C.~{De~Michele}, S.~Gabrielli, P.~Tartaglia and F.~Sciortino, \emph{J. Phys.
  Chem. B}, 2006, \textbf{110}, 8064\relax
\mciteBstWouldAddEndPuncttrue
\mciteSetBstMidEndSepPunct{\mcitedefaultmidpunct}
{\mcitedefaultendpunct}{\mcitedefaultseppunct}\relax
\EndOfBibitem
\bibitem[{De~Michele} \emph{et~al.}(2006){De~Michele}, Tartaglia, and
  Sciortino]{DeMicheleSilica06}
C.~{De~Michele}, P.~Tartaglia and F.~Sciortino, \emph{J. Chem. Phys.}, 2006,
  \textbf{125}, 204710\relax
\mciteBstWouldAddEndPuncttrue
\mciteSetBstMidEndSepPunct{\mcitedefaultmidpunct}
{\mcitedefaultendpunct}{\mcitedefaultseppunct}\relax
\EndOfBibitem
\bibitem[Zaccarelli \emph{et~al.}(2006)Zaccarelli, Saika-Voivod, Buldyrev,
  Moreno, Tartaglia, and Sciortino]{JCPnmax06}
E.~Zaccarelli, I.~Saika-Voivod, S.~V. Buldyrev, A.~J. Moreno, P.~Tartaglia and
  F.~Sciortino, \emph{J. Chem. Phys.}, 2006, \textbf{124}, 124908\relax
\mciteBstWouldAddEndPuncttrue
\mciteSetBstMidEndSepPunct{\mcitedefaultmidpunct}
{\mcitedefaultendpunct}{\mcitedefaultseppunct}\relax
\EndOfBibitem
\bibitem[Gado and Kob(2005)]{edgkob_epl05}
E.~D. Gado and W.~Kob, \emph{Europhysics Letters}, 2005, \textbf{72},
  1032\relax
\mciteBstWouldAddEndPuncttrue
\mciteSetBstMidEndSepPunct{\mcitedefaultmidpunct}
{\mcitedefaultendpunct}{\mcitedefaultseppunct}\relax
\EndOfBibitem
\bibitem[Gado and Kob(2007)]{edgkob_prl07}
E.~D. Gado and W.~Kob, \emph{Physical Review Letters}, 2007, \textbf{98},
  028303\relax
\mciteBstWouldAddEndPuncttrue
\mciteSetBstMidEndSepPunct{\mcitedefaultmidpunct}
{\mcitedefaultendpunct}{\mcitedefaultseppunct}\relax
\EndOfBibitem
\bibitem[Gado and Kob(2010)]{edgkob_sm10}
E.~D. Gado and W.~Kob, \emph{Soft Matter}, 2010, \textbf{6}, 1547\relax
\mciteBstWouldAddEndPuncttrue
\mciteSetBstMidEndSepPunct{\mcitedefaultmidpunct}
{\mcitedefaultendpunct}{\mcitedefaultseppunct}\relax
\EndOfBibitem
\bibitem[Foffi \emph{et~al.}(2005)Foffi, Michele, Sciortino, and
  Tartaglia]{fofficriJCP05}
G.~Foffi, C.~D. Michele, F.~Sciortino and P.~Tartaglia, \emph{J. Chem. Phys.},
  2005, \textbf{122}, 224903\relax
\mciteBstWouldAddEndPuncttrue
\mciteSetBstMidEndSepPunct{\mcitedefaultmidpunct}
{\mcitedefaultendpunct}{\mcitedefaultseppunct}\relax
\EndOfBibitem
\bibitem[Gado \emph{et~al.}(2003)Gado, Fierro, de~Arcangelis, and
  Coniglio]{ema_epl03}
E.~D. Gado, A.~Fierro, L.~de~Arcangelis and A.~Coniglio, \emph{Europhys.
  Lett.}, 2003, \textbf{63}, 1\relax
\mciteBstWouldAddEndPuncttrue
\mciteSetBstMidEndSepPunct{\mcitedefaultmidpunct}
{\mcitedefaultendpunct}{\mcitedefaultseppunct}\relax
\EndOfBibitem
\bibitem[Puertas \emph{et~al.}(2002)Puertas, Fuchs, and Cates]{puertas02}
A.~Puertas, M.~Fuchs and M.~Cates, \emph{Phys. Rev. Lett.}, 2002, \textbf{88},
  098301\relax
\mciteBstWouldAddEndPuncttrue
\mciteSetBstMidEndSepPunct{\mcitedefaultmidpunct}
{\mcitedefaultendpunct}{\mcitedefaultseppunct}\relax
\EndOfBibitem
\bibitem[Fierro \emph{et~al.}(2008)Fierro, Gado, de~Candia, and
  Coniglio]{af_jstat08}
A.~Fierro, E.~D. Gado, A.~de~Candia and A.~Coniglio, \emph{Journal of
  Statistical Mechanics-Theory and Experiment}, 2008,  L04002\relax
\mciteBstWouldAddEndPuncttrue
\mciteSetBstMidEndSepPunct{\mcitedefaultmidpunct}
{\mcitedefaultendpunct}{\mcitedefaultseppunct}\relax
\EndOfBibitem
\bibitem[Ottochian \emph{et~al.}(2009)Ottochian, {De~Michele}, and
  Leporini]{JCPwoly2009}
A.~Ottochian, C.~{De~Michele} and D.~Leporini, \emph{J. Chem. Phys.}, 2009,
  \textbf{131}, 224517\relax
\mciteBstWouldAddEndPuncttrue
\mciteSetBstMidEndSepPunct{\mcitedefaultmidpunct}
{\mcitedefaultendpunct}{\mcitedefaultseppunct}\relax
\EndOfBibitem
\bibitem[Widmer-Cooper and Harrowell(2006)]{Harrowell06}
A.~Widmer-Cooper and P.~Harrowell, \emph{Phys. Rev. Lett.}, 2006, \textbf{96},
  185701(4)\relax
\mciteBstWouldAddEndPuncttrue
\mciteSetBstMidEndSepPunct{\mcitedefaultmidpunct}
{\mcitedefaultendpunct}{\mcitedefaultseppunct}\relax
\EndOfBibitem
\bibitem[Glotzer and Vogel(2004)]{VogGlotz04}
S.~C. Glotzer and M.~. Vogel, \emph{Phys. Rev. E}, 2004, \textbf{70},
  061504\relax
\mciteBstWouldAddEndPuncttrue
\mciteSetBstMidEndSepPunct{\mcitedefaultmidpunct}
{\mcitedefaultendpunct}{\mcitedefaultseppunct}\relax
\EndOfBibitem
\bibitem[Buchenau and Zorn(1992)]{buchenau}
U.~Buchenau and R.~Zorn, \emph{Europhys. Lett.}, 1992, \textbf{18},
  523--528\relax
\mciteBstWouldAddEndPuncttrue
\mciteSetBstMidEndSepPunct{\mcitedefaultmidpunct}
{\mcitedefaultendpunct}{\mcitedefaultseppunct}\relax
\EndOfBibitem
\bibitem[Cornicchi \emph{et~al.}(2005)Cornicchi, Onori, and
  Paciaroni]{Paciaroni05}
E.~Cornicchi, G.~Onori and A.~Paciaroni, \emph{Phys. Rev. Lett.}, 2005,
  \textbf{95}, 158104\relax
\mciteBstWouldAddEndPuncttrue
\mciteSetBstMidEndSepPunct{\mcitedefaultmidpunct}
{\mcitedefaultendpunct}{\mcitedefaultseppunct}\relax
\EndOfBibitem
\bibitem[Magazu` \emph{et~al.}(2004)Magazu`, Maisano, and Migliardo]{Magazu04}
S.~Magazu`, G.~Maisano and F.~Migliardo, \emph{J.Chem.Phys.}, 2004,
  \textbf{121}, 8911--8915\relax
\mciteBstWouldAddEndPuncttrue
\mciteSetBstMidEndSepPunct{\mcitedefaultmidpunct}
{\mcitedefaultendpunct}{\mcitedefaultseppunct}\relax
\EndOfBibitem
\bibitem[Sokolov \emph{et~al.}(1993)Sokolov, R\"ossler, Kisliuk, and
  Quitmann]{SokolPRL}
A.~P. Sokolov, E.~R\"ossler, A.~Kisliuk and D.~Quitmann, \emph{Phys. Rev.
  Lett.}, 1993, \textbf{71}, 2062--2065\relax
\mciteBstWouldAddEndPuncttrue
\mciteSetBstMidEndSepPunct{\mcitedefaultmidpunct}
{\mcitedefaultendpunct}{\mcitedefaultseppunct}\relax
\EndOfBibitem
\bibitem[Sciortino and Kob(2001)]{KobFraPRLDW01}
F.~Sciortino and W.~Kob, \emph{Phys. Rev. Lett.}, 2001, \textbf{86},
  648--651\relax
\mciteBstWouldAddEndPuncttrue
\mciteSetBstMidEndSepPunct{\mcitedefaultmidpunct}
{\mcitedefaultendpunct}{\mcitedefaultseppunct}\relax
\EndOfBibitem
\bibitem[B\"assler(1987)]{Bassler87}
H.~B\"assler, \emph{Phys. Rev. Lett.}, 1987, \textbf{58}, 767--770\relax
\mciteBstWouldAddEndPuncttrue
\mciteSetBstMidEndSepPunct{\mcitedefaultmidpunct}
{\mcitedefaultendpunct}{\mcitedefaultseppunct}\relax
\EndOfBibitem
\bibitem[Ferry \emph{et~al.}(1953)Ferry, Grandine, and Fitzgerald]{FerryEtal53}
J.~D. Ferry, L.~D.~J. Grandine and E.~R. Fitzgerald, \emph{J. Appl. Phys.},
  1953, \textbf{24}, 911--916\relax
\mciteBstWouldAddEndPuncttrue
\mciteSetBstMidEndSepPunct{\mcitedefaultmidpunct}
{\mcitedefaultendpunct}{\mcitedefaultseppunct}\relax
\EndOfBibitem
\bibitem[Garrahan and Chandler(2003)]{EastArrow}
J.~P. Garrahan and D.~Chandler, \emph{Proc. Natl. Acad. Sci.}, 2003,
  \textbf{100}, 9710\relax
\mciteBstWouldAddEndPuncttrue
\mciteSetBstMidEndSepPunct{\mcitedefaultmidpunct}
{\mcitedefaultendpunct}{\mcitedefaultseppunct}\relax
\EndOfBibitem
\bibitem[Monthus and Bouchaud(1996)]{MonthBouch96}
C.~Monthus and J.-P. Bouchaud, \emph{J. Phys. A: Math. Gen.}, 1996,
  \textbf{29}, 3847--3869\relax
\mciteBstWouldAddEndPuncttrue
\mciteSetBstMidEndSepPunct{\mcitedefaultmidpunct}
{\mcitedefaultendpunct}{\mcitedefaultseppunct}\relax
\EndOfBibitem
\bibitem[Laurati \emph{et~al.}(2009)Laurati, Petekidis, Koumakis, Cardinaux,
  Schofield, Brader, Fuchs, and Egelhaaf]{exp2}
M.~Laurati, G.~Petekidis, N.~Koumakis, F.~Cardinaux, A.~B. Schofield, J.~M.
  Brader, M.~Fuchs and S.~U. Egelhaaf, \emph{J. Chem. Phys.}, 2009,
  \textbf{130}, 134907\relax
\mciteBstWouldAddEndPuncttrue
\mciteSetBstMidEndSepPunct{\mcitedefaultmidpunct}
{\mcitedefaultendpunct}{\mcitedefaultseppunct}\relax
\EndOfBibitem
\bibitem[Dinsmore \emph{et~al.}(2006)Dinsmore, Prasad, Wong, and
  Weitz]{dinsmore_prl06}
A.~D. Dinsmore, V.~Prasad, I.~Y. Wong and D.~A. Weitz, \emph{Phys. Rev. Lett.},
  2006, \textbf{96}, 185502\relax
\mciteBstWouldAddEndPuncttrue
\mciteSetBstMidEndSepPunct{\mcitedefaultmidpunct}
{\mcitedefaultendpunct}{\mcitedefaultseppunct}\relax
\EndOfBibitem
\bibitem[Dibble \emph{et~al.}(2008)Dibble, Kogan, and Solomon]{solomon}
C.~J. Dibble, M.~Kogan and M.~J. Solomon, \emph{Phys. Rev. E}, 2008,
  \textbf{77}, 050401\relax
\mciteBstWouldAddEndPuncttrue
\mciteSetBstMidEndSepPunct{\mcitedefaultmidpunct}
{\mcitedefaultendpunct}{\mcitedefaultseppunct}\relax
\EndOfBibitem
\bibitem[Ohtsuka \emph{et~al.}(2008)Ohtsuka, Royall, and Tanaka]{royall}
T.~Ohtsuka, C.~P. Royall and H.~Tanaka, \emph{Europhys. Lett.}, 2008,
  \textbf{84}, 46002\relax
\mciteBstWouldAddEndPuncttrue
\mciteSetBstMidEndSepPunct{\mcitedefaultmidpunct}
{\mcitedefaultendpunct}{\mcitedefaultseppunct}\relax
\EndOfBibitem
\bibitem[Miller \emph{et~al.}(2009)Miller, Blaak, Lumb, and Hansen]{miller}
M.~A. Miller, R.~Blaak, C.~N. Lumb and J.-P. Hansen, \emph{J. Chem. Phys.},
  2009, \textbf{130}, 114507\relax
\mciteBstWouldAddEndPuncttrue
\mciteSetBstMidEndSepPunct{\mcitedefaultmidpunct}
{\mcitedefaultendpunct}{\mcitedefaultseppunct}\relax
\EndOfBibitem
\bibitem[Saw \emph{et~al.}(2009)Saw, Ellegaard, Kob, and Sastry]{kob_sastry_09}
S.~Saw, N.~L. Ellegaard, W.~Kob and S.~Sastry, \emph{Phys. Rev. Lett.}, 2009,
  \textbf{103}, 248305\relax
\mciteBstWouldAddEndPuncttrue
\mciteSetBstMidEndSepPunct{\mcitedefaultmidpunct}
{\mcitedefaultendpunct}{\mcitedefaultseppunct}\relax
\EndOfBibitem
\bibitem[Allen and Tildesley(1989)]{md}
M.~P. Allen and D.~Tildesley, \emph{Computer simulation of liquids}, Clarendon
  Press, Oxford, 1989\relax
\mciteBstWouldAddEndPuncttrue
\mciteSetBstMidEndSepPunct{\mcitedefaultmidpunct}
{\mcitedefaultendpunct}{\mcitedefaultseppunct}\relax
\EndOfBibitem
\bibitem[Rubinstein and Colby(2003)]{rubinstein}
M.~Rubinstein and R.~H. Colby, \emph{Polymer Physics}, Oxford University Press,
  Oxford, 2003\relax
\mciteBstWouldAddEndPuncttrue
\mciteSetBstMidEndSepPunct{\mcitedefaultmidpunct}
{\mcitedefaultendpunct}{\mcitedefaultseppunct}\relax
\EndOfBibitem
\bibitem[Gado and Kob(2008)]{edgkob_jnnfm08}
E.~D. Gado and W.~Kob, \emph{Journal of Non-Newtonian Fluid Mechanics}, 2008,
  \textbf{149}, 28\relax
\mciteBstWouldAddEndPuncttrue
\mciteSetBstMidEndSepPunct{\mcitedefaultmidpunct}
{\mcitedefaultendpunct}{\mcitedefaultseppunct}\relax
\EndOfBibitem
\end{mcitethebibliography}
\bibliographystyle{rsc} 
}

\end{document}